\documentclass[12pt]{article}
\usepackage{authblk}

\usepackage{soul} 
\usepackage{titlesec}

\titleformat*{\section}{\normalsize\bfseries}
\titleformat*{\subsection}{\normalsize\bfseries}

\newcommand{\Add}[1]{\textcolor{black}{#1}}
\newcommand{\Erase}[1]{\iffalse#1\fi} 

\usepackage[utf8]{inputenc}
\usepackage{amsmath}
\usepackage{graphicx}
\usepackage{hyperref}
\hypersetup{
    colorlinks=true,
    linkcolor=blue,
    filecolor=blue,      
    urlcolor=blue,
    citecolor=blue,
}

\title{300~GHz wave with attosecond-level timing noise}
\author[1]{Tomohiro Tetsumoto}
\author[2]{Tadao Nagatsuma}
\author[3]{Martin E. Fermann}
\author[4]{Gabriele Navickaite}
\author[4]{Michael Geiselmann}
\author[1,*]{Antoine Rolland}
\affil[1]{IMRA America Inc., Boulder Research Labs, 1551 South Sunset St, Suite C, Longmont, Colorado 80501, USA}
\affil[2]{Graduate School of Engineering Science, Osaka University, 1-3 Machikaneyama, Toyonaka, Osaka 560-8531, Japan}
\affil[3]{IMRA America Inc., 1044 Woodridge Ave, Ann Arbor, Michigan 48105, USA}
\affil[4]{LIGENTEC SA, EPFL Innovation Park L, Chemin de la Dent d’Oche 1B, Switzerland CH-1024 Ecublens, Switzerland}
\affil[*]{Corresponding author: arolland@imra.com}
\date{}                     
\usepackage[sorting=none, refsegment=section, style=numeric, bibstyle=nature]{biblatex}
\usepackage{textcomp}
\usepackage[left=1.6cm, right=1.6cm, top=2.0cm, bottom=2.0cm]{geometry}
\usepackage{lscape}

\begin{document}
\maketitle
\textbf{Optical frequency division (OFD) via optical frequency combs (OFC) has enabled a leap in microwave metrology leading to noise performance never explored before. Extending the method to millimeter-wave (mmW) and terahertz (THz)-wave domain is of great interest. Dissipative Kerr solitons (DKSs) in integrated photonic chips offer a unique feature of delivering OFCs with ultrahigh repetition rates from 10~GHz to 1~THz making them relevant gears to perform OFD in the mmW and THz-wave domain. We experimentally demonstrate OFD of an optically-carried 3.6~THz reference down to 300~GHz through a DKS, photodetected with an ultrafast uni-traveling-carrier photodiode (UTC-PD). A new measurement system, based on the characterization of a microwave reference phase-locked to the 300~GHz signal under test, yields attosecond-level timing noise sensitivity, leveraging conventional technical limitations. This work places DKSs as a leading technology in the mmW and THz-wave field promising breakthroughs in fundamental and civilian applications.}
\newline

\begin{refsection}
\noindent
\Add{Demands for spectrally pure mmWs (30-300~GHz) and THz-waves (0.1-10~THz) are growing in a number of modern-world applications~\cite{nagatsuma2015millimeter, koenig2013wireless, dhillon20172017}. Tireless efforts to expand data capacity continues in the field of wireless communication for 5G (frequency of $<90$~GHz) and 6G (frequency of $<$~multi-THz) deployment~\cite{dang2020should, Zhang2019}, where low phase noise oscillators help increasing the number of bits per symbol in phase modulation schemes while keeping bit-error-ratio minimized. Precise resolving of rotational spectra of molecules \Add{and chip-scale molecular clocks} in the frequency range require a narrow-linewidth oscillator~\cite{wang2018chip}. Moreover, scientific applications including radio astronomy~\cite{de2010herschel,brunken2014h} and laser driven THz electron accelerators~\cite{nanni2015terahertz,zhang2018segmented}, military~\cite{Grajal2017} and civilian applications such as remote sensing~\cite{RemoteSensing2017} and automotive radar~\cite{Patole7870764} will ultimately benefit from state-of-the-art instrument-related noise and stability.}
\Add{The ultimate in} spectrally pure and accurate electromagnetic waves with extreme performances \Erase{lie}\Add{is available} in the optical domain~\cite{Matei2017, Huntemann2016, mcgrew2018atomic, brewer2019al+}. Challenges encountered by direct microwave generation in terms of bandwidth, spectral purity and stability have been overcome by the application of novel technologies based on the generation of coherent and low phase noise optical signals, such as, optoelectronic oscillators~\cite{Yao:96}, electro-optic (EO) frequency division~\cite{li2014electro}, Brillouin oscillators~\cite{Li:19}, OFD~\cite{fortier2011generation}, and high-\textit{Q} microresonator oscillators~\cite{li2012low,liang2015high,lucas2020ultralow}.
\Add{In principle, similar methods can also be adapted to the mmW and THz-wave domain facilitated through ultrafast optoelectronic devices, recently made available as results of extensive research in relevant photonics and electronics areas~\cite{ishibashi2014unitraveling, wang2019room, ummethala2019thz}.}

\Add{OFD of a continuous wave (CW) laser through an OFC has enabled the generation of a microwave signal with unprecedented absolute phase noise level and fractional frequency stability~\cite{xie2017photonic, nakamura2020coherent}.}
In order to exploit the same mechanism to create mmW and THz-wave oscillators, several challenges need to be addressed. OFD to the mmW domain using bulky OFCs is no longer relevant as their relatively low repetition rates make photodetection in, \Add{or EO multiplication~\cite{fortier2016optically} to the mmW domain rather inefficient.}
On the other hand, DKS combs with multi-hundreds of GHz repetition rates based on \Add{integrated} microresonators~\cite{kippenberg2018dissipative} have an apropos potential in generating mmWs and THz-waves with direct photomixing techniques.
\Add{DKS combs have successfully facilitated frequency synthesis in the microwave~\cite{liu2020photonic,wang2020vernier,xu2019broadband}, optical~\cite{spencer2018optical}, and even mmW domain at 300~GHz recently~\cite{zhang2019terahertz, tetsumoto2020300}}.
While those approaches serve as important alternatives to microwave multipliers or bulky OFCs, they rely on a microwave reference whose phase fluctuations are multiplied up to the frequency range of interest. 
One fundamental challenge, for the use of a DKS comb in an OFD scheme, is a demonstration of conventional self-referencing of the optical pulse trains due to its low pulse energy in exchange for the high repetition rate. This makes frequency doubling, an essential element of self-referencing, non-trivial in particular when one of the desired outcomes is integration of the concept.
In contrast, using two low noise CW lasers with a large frequency separation for phase-locking the two degrees of freedom of an OFC (i.e., a repetition rate frequency $f_\mathrm{rep}$ and a carrier envelop offset frequency $f_\mathrm{ceo}$) removes the need of self-referencing at the expense of a smaller frequency division ratio~\cite{li2014electro, swann2011microwave}.

\Add{Here, we demonstrate OFD of a DKS comb by employing two low noise Brillouin Stokes waves for the first time in the mmW domain. The small dimension of a microresonator with a diameter of about 150~\textmu m enables a DKS comb with an ultrahigh repetition rate of 300~GHz and efficient generation of a mmW through photomixing at a UTC-PD. To characterize the realized ultra-low phase noise of a generated 300~GHz carrier, we develop a new measurement system based on a delay-line interferometer and EO combs driven by a microwave oscillator phase-locked to a stabilized DKS comb. 
The measurement system enables accessing record phase noise of $-100$~dBc/Hz (8~as.Hz$^{-1/2}$ in timing noise) at a Fourier frequency of 10~kHz for the 300~GHz carrier, which is one-order of magnitude lower than measured 300~GHz wave noise based on any other technologies. Surprisingly, the obtained value is still limited by the measurement and smaller repetition rate noise of $-108$~dBc/Hz (3~as.Hz$^{-1/2}$ in timing noise) at 10~kHz is evaluated indirectly.
The implementation of DKS combs as demonstrated here opens the path towards full integration of precision mmW and THz-wave comb systems, and the establishment of DKSs in innumerable applications in science and technology.}

\section*{Results}
\section*{\Add{Phase noise reduction through optical frequency division}}

\begin{figure*}[!ht]
    \centering
    \includegraphics[width=\linewidth]{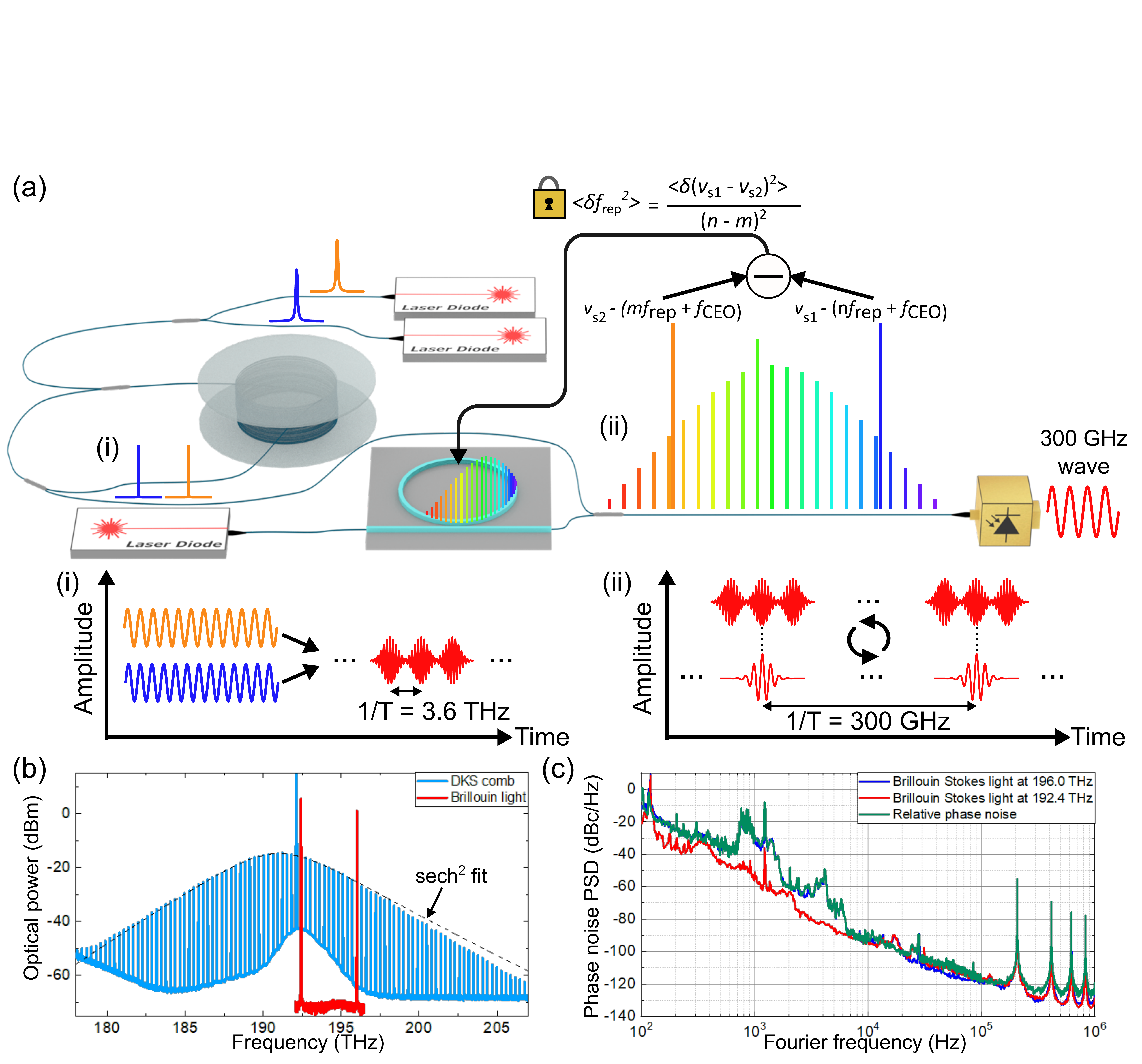}
    \caption{\footnotesize{\textbf{Coherent optical frequency division of a 3.6~THz reference down to a photodetected 300~GHz optical pulse train. }(\textbf{a}) Conceptual sketch of the 300~GHz oscillator. An optically-carried 3.6~THz reference (i), generated from two spatially overlapped Stokes waves from stimulated Brillouin scattering in a 75-m-long fiber ring cavity, is ruling the timing of the repetition rate frequency of a DKS comb at 300~GHz~(ii). This type of synchronization is known as optical frequency division. (\textbf{b}) Optical spectrum \Erase{combining the 3.6~THz dual-wavelength}\Add{of Brillouin} Stokes waves \Erase{with}\Add{and} the output of the DKS comb. The envelop of the DKS comb follows sech$^2$ function fit overall. The observed mismatch at short and long wavelengths is mainly due to operation bandwidth of optical components. (\textbf{c}) \Erase{Power spectral density}PSD of phase noise of the two Stokes wave at 192.4~THz (red curve) and 196.0~THz (blue curve) measured with a 1-km-optical-delay based self-heterodyne interferometer. The green curve exhibits the differential phase noise between the two Stokes waves (i.e., the phase noise of the optically-carried 3.6~THz reference) measured with a TWDI with a 1~km fiber delay.}}
    \label{fig1}
\end{figure*}

\noindent
The working principle of the ultra-low noise 300~GHz oscillator is \Add{schematically} depicted in Fig.~\ref{fig1}(a). An optically-carried 3.6~THz reference is generated \Add{through stimulated Brillouin scattering in} a 75-m-long fiber ring cavity pumped by two CW lasers with frequency separation of 3.6~THz: frequencies of $\nu_\mathrm{s1}=196.0$~THz and $\nu_\mathrm{s2}=192.4$~THz (see Fig.~\ref{fig1}(b)).
The 3.6~THz reference beats with a DKS comb with a 300~GHz repetition rate pumped at 192.1~THz. Like any other OFCs, DKS combs are governed by two quantities known as the repetition rate frequency $f_\mathrm{rep}$ and the carrier envelop offset frequency $f_\mathrm{ceo}$. When both optical references $\nu_\mathrm{s1}$ and $\nu_\mathrm{s2}$ beat on a photodetector with the DKS comb, two beatnotes are generated at RF frequencies of:

\begin{equation}
     f_\mathrm{b1} =  \nu_\mathrm{s1}- (n f_\mathrm{rep}+f_\mathrm{ceo}),
    \label{equ1}
\end{equation}
\begin{equation}
     f_\mathrm{b2} =  \nu_\mathrm{s2}- (m f_\mathrm{rep}+f_\mathrm{ceo}).
    \label{equ2}
\end{equation}
\noindent
Taking the difference of $f_\mathrm{b1}$ and $f_\mathrm{b2}$ removes the phase fluctuations of the carrier envelop offset frequency $f_\mathrm{ceo}$. Doing so results in an intermediate frequency $f_\mathrm{x}$ whose phase fluctuations can be written as:

\begin{equation}
     \delta f_\mathrm{x} = \delta (\nu_\mathrm{s1}-\nu_\mathrm{s2})-(n-m) \delta f_\mathrm{rep}
    \label{equ3}
\end{equation}
\noindent
Nullifying the phase difference of $f_\mathrm{x}$ and a RF local oscillator, such as $\delta f_x=0$ (as \Add{phase fluctuation of the RF oscillator} can be ignored), leads to an expression for the phase noise at 300~GHz:

\begin{equation}
    <\delta f_\mathrm{rep}^2> =\frac{<\delta (\nu_\mathrm{s1}-\nu_\mathrm{s2})^2>}{(n-m)^2}.
    \label{equ4}
\end{equation}
\noindent
\Erase{leading to}{This gives} a phase noise reduction \Erase{corresponding to}\Add{by} the frequency ratio of the 3.6~THz reference and the repetition rate frequency 300~GHz\Erase{.}\Add{, which corresponds to $20 \times \log{\frac{3.6~THz}{300~GHz}}=21.6$~dB} \Erase{I}\Add{i}n terms of power spectral density \Add{(PSD)} of phase noise. 
\Add{Phase noise PSD} of both optical lines have been evaluated \Add{individually} using a self-heretodyne interferometer comprising an optical fiber delay of 1~km. Additionally and more importantly, we have measured the optically-carried 3.6~THz reference with a two-wavelength delayed self-heterodyne interferometer (TWDI)~\cite{kuse2017electro}. Indeed, the relative phase noise of the two Stokes waves is the one divided down to 300~GHz \Add{in the OFD experiment}. \Add{Phase noise PSD} carried by the two pump lasers is entirely decorrelated. So, phase noise of the two Stokes waves, generated through stimulated Brillouin scattering, is directly governed by the phase noise of the pumps \Add{but} with reduced noise level thanks to acoustic damping and cavity feedback mechanism \Add{(see supplementary information for details)}.
Phase noise of the two optical Stokes waves (red and blue curves) as well as their differential phase noise at 3.6~THz (green cruve) are shown in Fig.~\ref{fig1}(c). 
The phase noise of both Stokes waves is recorded with levels of -95~dBc/Hz at 10~kHz and -120~dBc/Hz at 100~kHz, respectively. Differential phase noise of the two Stokes waves, leading to an optically-carried reference at 3.6~THz, is ruled by the Stokes wave with higher noise. As a result, the 3.6~THz reference exhibits the same phase noise \Add{level as the 196.0~THz laser at Fourier frequencies from 100~Hz to 10~kHz and the same phase noise level as both lasers} for Fourier frequencies from 10~kHz to 100~kHz. 

\begin{figure*}
    \centering
    \includegraphics[width=14cm]{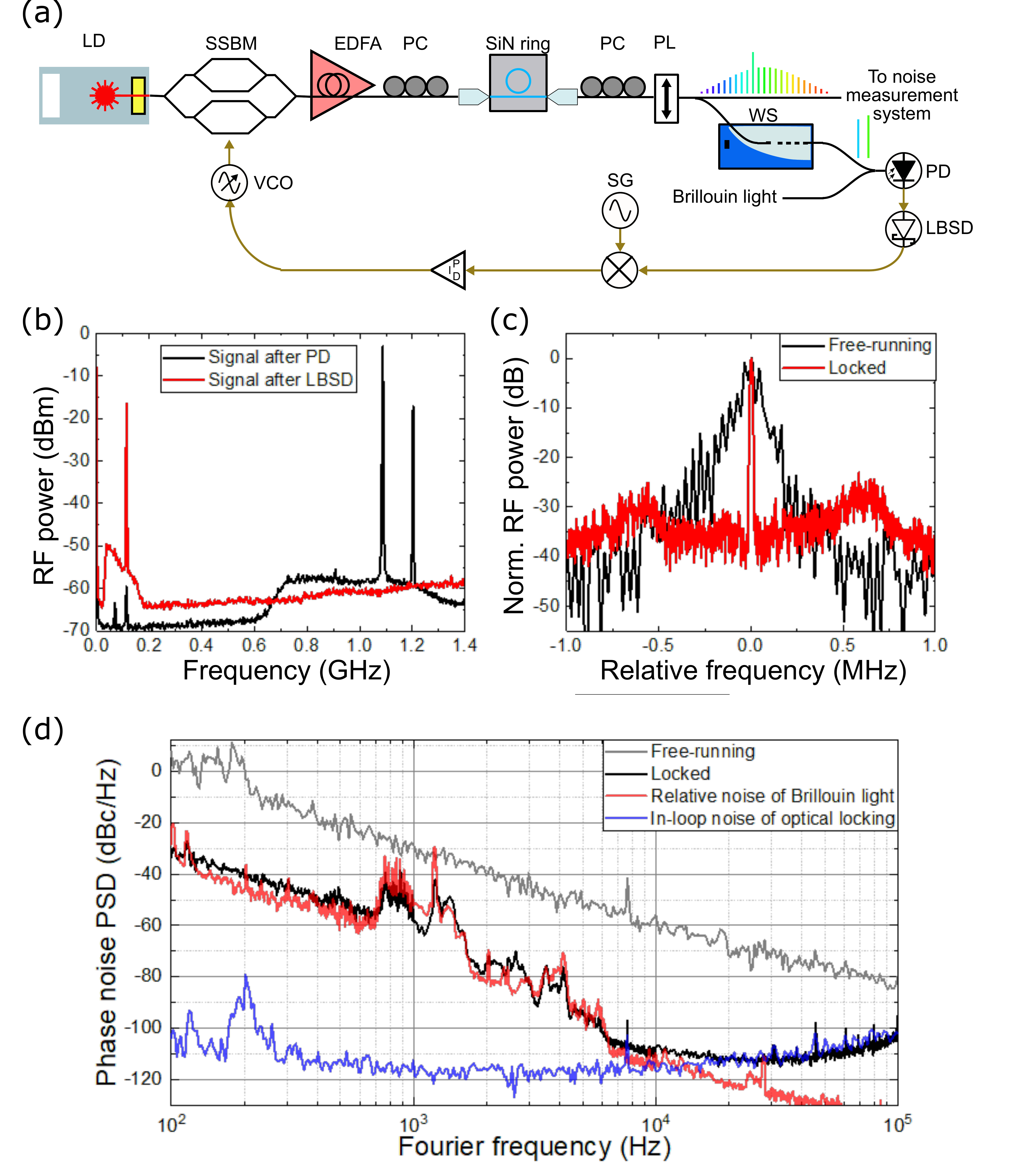}
    \caption{\footnotesize{\textbf{Optical phase locked loop for synchronization of the 300~GHz pulse train to the 3.6~THz reference.} (\textbf{a}) Experimental setup describing the phase fluctuation control of the DKS comb 300~GHz repetition rate. EDFA: erbium doped fiber amplifier. PC: polarization controller. PL: polarizer. VCO: voltage controlled oscillator. See text for details.
    (\textbf{b}) RF spectra of the two RF beatnotes (black curve) at the output of the photodetector, and the intermediate frequency (red curve) at the output of the Schottky diode. (\textbf{c}) RF spectra of the intermediate frequency $f_\mathrm{x}$. Resolution bandwidth is 10~kHz. Y axis is normalized so that maximum power takes 0~dB. (\textbf{d}) \Erase{Power spectral density}\Add{PSD} of phase noise of the free-running DKS comb repetition rate at 300~GHz (grey curve), phase-locked DKS comb repetition rate at 300~GHz (black curve), 3.6~THz reference sacled to 300~GHz assuming perfect divsion (red curve), and optical in-loop servo error scaled to 300~GHz (blue curve).}}
    \label{fig2}
\end{figure*}

To divide down the reference at 3.6~THz to 300~GHz, we have implemented the experimental setup depicted in Fig.~\ref{fig2}(a). A 300~GHz DKS comb is initiated by a fast frequency sweep of the CW pump light with a single side band modulator (SSBM) (see supplementary information for details). The DKS comb is split into two arms for measurement purpose (upper arm) and servo control purpose (lower arm), respectively. The lower arm is sent to a waveshaper (WS) which spectrally selects two comb modes of interest to beat with the 3.6~THz dual-wavelength reference at a photodiode (PD) (the generated two beatnotes are at around 1.1~GHz on black curve in Fig.~\ref{fig2}(b)). The output of the PD is sent directly to a low-barrier Schottky diode (LBSD), used as a microwave amplitude detector, that generate an intermediate RF frequency $f_\mathrm{x}$ whose noise corresponds to Eq.~\ref{equ3} (shown at around 100~MHz on red curve in Fig.~\ref{fig2}(b)). The RF signal $f_\mathrm{x}$ is mixed down to direct current (DC) with a signal generator (SG), generating an error signal, and fed back to \Add{the frequency modulation input of} the SSBM through a proportional-integral-derivative (PID) loop filter~\cite{kuse2019control}.
Enlarged RF spectra of the intermediate frequency $f_\mathrm{x}$ are shown in Fig.~\ref{fig2}(c). We can observe that the phase locked loop is effective by analyzing the in-loop RF signal in free-running (black curve on Fig.~\ref{fig2}(c)) and locked operation \Add{showing a coherent peak} (red curve on Fig.~\ref{fig2}(c)). The locked in-loop RF signal contains important servo bumps and contributes to the residual phase error due to the phase locked loop. However, one strong advantage of optical frequency division is that the residual phase error $\delta \varepsilon$ is also divided down by the same ratio as the reference frequency is divided by. Therefore, Eq.~\ref{equ4} needs be re-written as follows:
\begin{equation}
    <\delta f_\mathrm{rep}^2> =\frac{<\delta (\nu_\mathrm{s1}-\nu_\mathrm{s2})^2>+<\delta \varepsilon ^2>}{(n-m)^2}.
    \label{equ5}
\end{equation}

\noindent
The contribution of the fluctuation of the SG used to down convert $f_\mathrm{x}$ to DC can be ignored at this stage as it is comparing a 100~MHz carrier with a 300~GHz carrier. As a preliminary evaluation of the phase noise of the 300~GHz repetition rate, we have carried out a measurement of the phase noise following the method described in a reference~\cite{kuse2017electro} by using a TWDI with a 2-km-long optical fiber delay. Fig.~\ref{fig2}(d) highlights the measured phase noise of the 300~GHz repetition rate and other signals of interest that are contained in Eq.~\ref{equ5}. We have scaled the 3.6~THz reference phase noise down to 300~GHz (red curve on Fig.~\ref{fig2}(d)) as well as the residual phase error (blue curve on Fig.~\ref{fig2}(d)). According to Eq.~\ref{equ5}, the resulting phase noise of the repetition rate is the quadratic sum of the 3.6~THz phase noise and the residual phase error\Erase{(i.e., the convolution of those two fluctuation signals)}. This is confirmed by the measurement and exhibited on the black curve of Fig.~\ref{fig2}(d).The repetition rate phase noise faithfully follows the 3.6~THz reference, for Fourier frequencies lower than 10~kHz, until the phase fluctuation is dominated by the residual phase error, for Fourier frequencies higher than 10~kHz. The out-of-loop phase noise of the 300~GHz repetition rate reaches $-108$~dBc/Hz at 10~kHz Fourier frequency~(about 50~dB lower than the free-running noise shown on the grey curve of Fig.~\ref{fig2}(d)), which represents record at 300~GHz (the lowest measured number is -90~dBc/Hz at 10~kHz to the best of our knowledge~\cite{Li:19}).

\section*{Generation and characterization of a 300~GHz signal}
\begin{figure*}
    \centering
    \includegraphics[width=\linewidth]{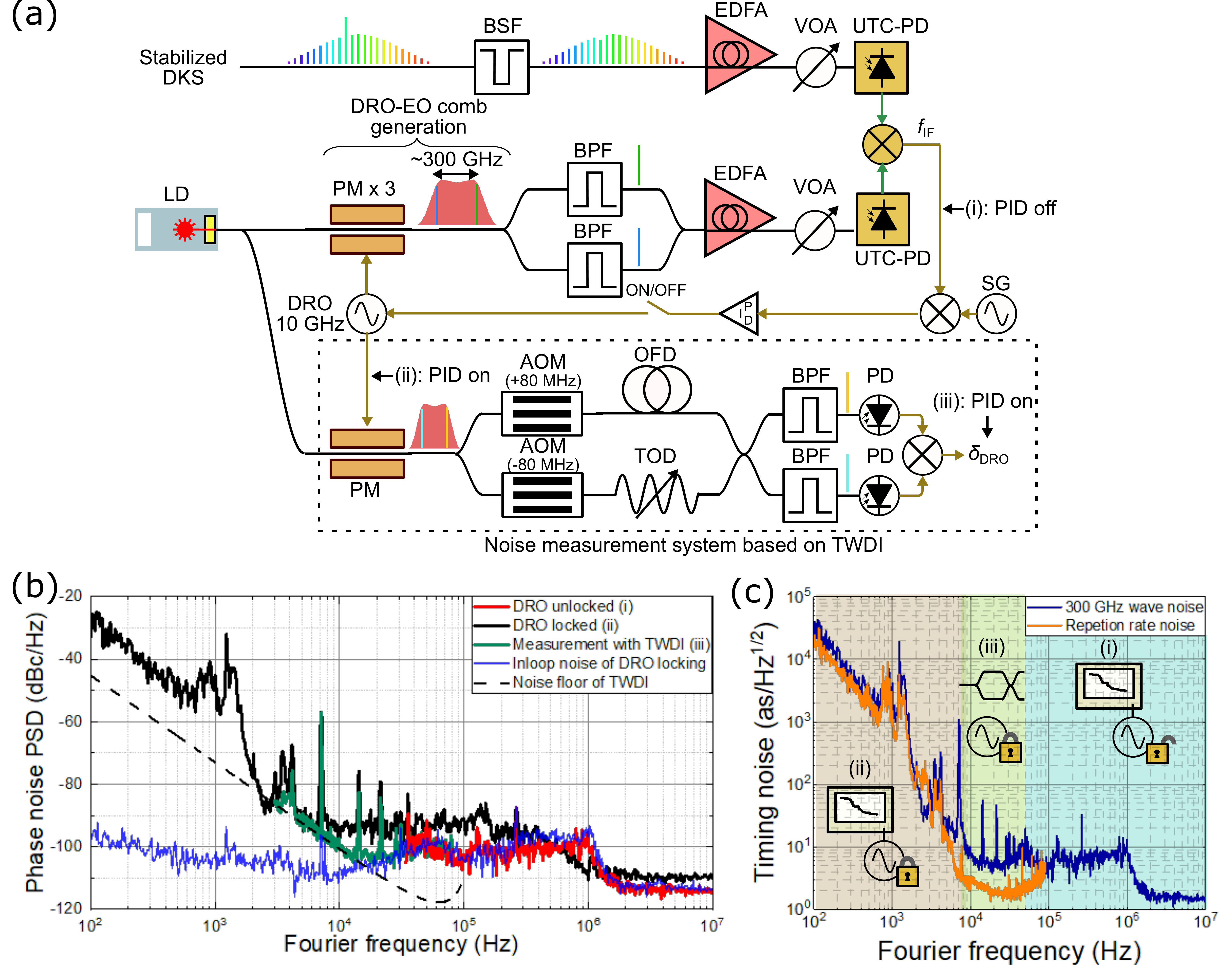}
    \caption{\footnotesize{ \textbf{Measurement methods and characterization of the spectral purity of the 300~GHz wave} (\textbf{a}) Experimental setup aiming at characterizing the \Erase{power spectral density}\Add{PSD} of phase noise of the 300~GHz wave. VOA: variable optical attenuator. BSF: band stop filter. BPF: band pass filter. PM: phase modulator. AOM: Acousto-optic modulator. OFD: optical fiber delay. TOD: tunable optical delay. See text for details. (\textbf{b}) \Erase{Power spectral density}\Add{PSD} of phase noise of the 300~GHz wave. The measurement is performed using three different methods. For Fourier frequencies of $>50$~kHz, the down-converted signal is directly measured with a phase noise analyzer (i). For Fourier frequencies of $<10$~kHz, the phase noise is obtained from the phase-locked DRO at 10~GHz measured with a phase noise analyzer and scaled to 300~GHz (ii). Measurement with a TWDI with a 2~km fiber delay is performed for Fourier frequencies between 10~kHz and 100~kHz (iii). (\textbf{c}) Timing noise of the 300~GHz wave (blue curve) and the 300~GHz repetition rate (orange curve). \Add{The measurement methods used for each Fourier frequency range of the blue curve are indicated by notations, icons and background colors.}}}
    \label{fig3}
\end{figure*}

\noindent
While the repetition rate phase noise is successfully phase locked to the 3.6~THz reference, no 300~GHz wave has yet been generated. \Add{For the generation and the characterization of a 300~GHz wave, we constructed a setup shown in Fig.~\ref{fig3}(a). The stabilized DKS comb with 300~GHz repetition rate illuminates a UTC-PD, which emits a mmW into a metal hollow core waveguide.}
\Add{Measuring expected low noise level in the mmW frequency range is far from trivial. Preparing a 300~GHz local oscillator through multiplication of a low noise microwave reference is difficult without adding extra noise besides intrinsic accumulated noise and inevitable conversion loss. Also, duplication of the oscillator under test will double the required equipment for the experiment.}
As an alternative, we have built a second mmW source based on EO multiplication of a 10~GHz dielectric resonant oscillator (DRO) \Add{phase locked to the stabilized DKS}\Erase{to 300~GHz by beating on a UTC-PD two spectrally-filtered EO comb sidebands separated by 300~GHz}.
\Add{As shown in Fig.~\ref{fig3}(a), two optical lines of the EO comb with frequency separation of 300~GHz are sampled and injected into another UTC-PD.}
The DKS-based 300~GHz (under test) is replicated to a baseband frequency $f_\mathrm{IF}$ by being down-converted with the DRO-EO-based 300~GHz through a fundamental frequency mixer at 300~GHz. The intermediate frequency $f_\mathrm{IF}$ carries the combination of the DRO noise and the DKS noise. 
To evaluate the noise of the DKS comb, \Erase{we phase lock the DRO to the DKS comb 300~GHz signal. }\Add{an error signal is generated from $f_\mathrm{IF}$ and applied to the DRO.}
The DRO at 10~GHz is then carrying the noise of the the DKS-based 300~GHz signal within the loop bandwidth of 400~kHz. Measuring phase noise in the microwave domain can be performed with an off-the-shelf phase noise analyzer.

Fig.~\ref{fig3}(b) summarizes the measured \Add{phase noise PSD} of signals of interest. \Add{We have introduced three measurement points to maximize the sensitivity.} The red curve is measured at the point (i) on the experimental setup \Add{in the condition that the DRO is not locked to the DKS}. \Erase{From previously reported data~, we know where the DRO phase noise multiplied to 300~GHz stands.} To lighten information on the graph we have only plotted the phase noise for Fourier frequency of 30~kHz to 10~MHz \Erase{which, we know, are dominated by the DKS comb noise}\Add{where, we know, the DKS noise is higher than the multiplied DRO noise}~\cite{kuse2017electro}. The limits on the measured phase noise are ruled by the residual phase error $\delta \varepsilon$ and the shot noise floor. The black curve is the phase noise of the DRO locked to the DKS comb at the point (ii) in the experimental setup. The measured noise of 10~GHz signal is scaled up to 300~GHz. \Add{The blue curve is the corresponding in-loop phase error at 300~GHz.} From Fourier frequencies of about 1~kHz to 10~kHz, the phase noise is measured to be governed by the 3.6~THz reference scaled to 300~GHz, as expected. The plateau at -95~dBc/Hz from 10~kHz to 100~kHz is not anticipated to come from either the residual phase error or the 3.6~THz reference. The phase noise plot beyond the loop bandwidth is irrelevant as it is merely the DRO noise multiplied up to 300~GHz. As we are limited by the noise floor of the instrument for Fourier frequencies from 10~kHz to 100~kHz, we have added a third point of measurement for this specific decade (labeled (iii) in Fig.~\ref{fig3}(a)). The phase locked DRO at 10~GHz is being measured with a real-time phase noise analyzer, with high sensitivity, based on an EO comb and a \Erase{1-km-long optical delay}\Add{TWDI with a 2~km fiber delay} (see supplementary information or a reference~\cite{kuse2017electro} for details). The black dashed curve on Fig.~\ref{fig3}(b) shows the measurement noise floor of the method. The measurement result is plotted from Fourier frequencies of 3~kHz to 100~kHz. Again, it becomes apparent that the measurement method itself limits the phase noise level of the DKS-based 300~GHz wave. Nevertheless, we measure a phase noise of $-100$~dBc/Hz at 10~kHz Fourier frequency, which \Add{is still one decade lower than the best value ever measured in the mmW domain at 300~GHz}~\cite{Li:19}. Above 20~kHz Fourier frequency, the phase noise is limited by the residual phase error of the DRO phase locked loop at 300~GHz.

Stitching the three points (i, ii, iii) of measurement in the Fourier frequency ranges of interest gives the timing noise of the 300~GHz wave under test (See Fig.~\ref{fig3}(c)). Timing noise is an effective figure of merit in order to compare one oscillator to another whose oscillation frequencies are different. Above 10~kHz Fourier frequency, the timing noise floor is lower than 10~as.Hz$^{-1/2}$ while the noise floor drops near 1~as.Hz$^{-1/2}$. We have superimposed the timing noise of the repetition rate at 300~GHz (i.e., before conversion to mmW, orange curve in Fig.~\ref{fig3}(c)). Timing noise at 10~kHz Fourier frequency is recorded at 3~as.Hz$^{-1/2}$. \Add{It is worth noting that this is about 12 times lower than that of a 20~GHz wave in a DKS based on a SiN microresonator (about $-110$~dBc/Hz at 10~kHz)~\cite{liu2020photonic} and same level as that for a 14~GHz wave based on a DKS from a crystalline resonator ($-135$~dBc/Hz at 10~kHz)~\cite{lucas2020ultralow}, respectively, despite the much higher carrier frequency.}

\begin{figure*}[!ht]
    \centering
    \includegraphics[width=400pt]{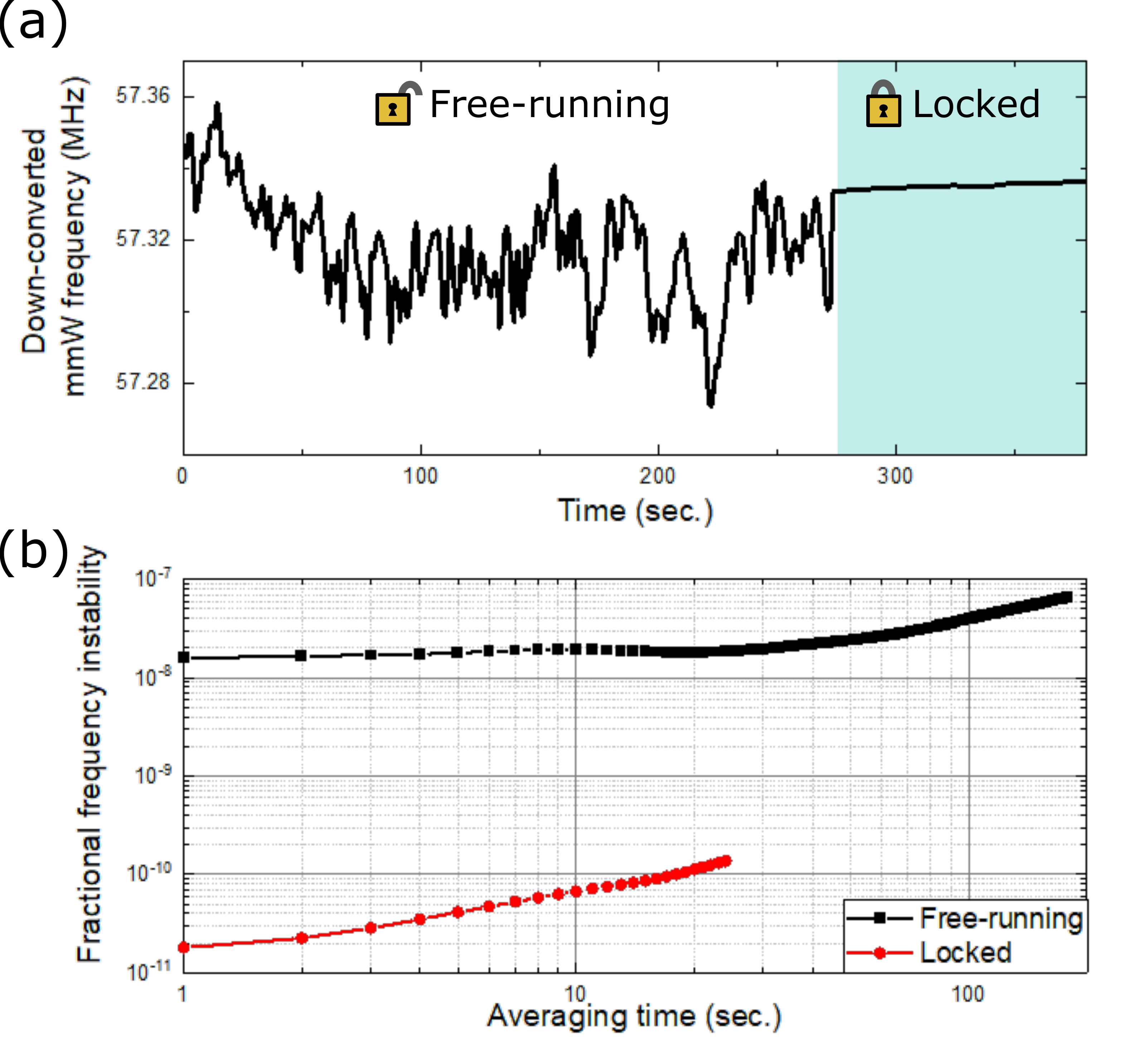}
    \caption{\footnotesize{\textbf{Characterization of the frequency instability of the 300~GHz wave} (\textbf{a}) Time trace of the beatnote between the DRO-EO-comb-based 300~GHz signal and the \Add{DKS-}based 300~GHz signal. (\textbf{b}) Fractional frequency stability in terms of modified Allan deviation of the \Add{DKS}-based 300~GHz signal in free-running (black curve) and locked to the 3.6~THz reference (red curve).}}
    \label{fig4}
\end{figure*}

To complete the characterization of an oscillator it is important to evaluate the mid- and long-term stability. With a frequency counter, we have recorded the intermediate frequency deviation, with 1 second gate time, (point (i) on Fig.~\ref{fig3}(a)) in open loop operation. The frequency counter is referenced with a 10~MHz output from a Rubidium clock. Fig.~\ref{fig4}(a) shows the intermediate frequency over time. Two areas are made distinct for free-running operation of the DKS comb and when the DKS comb is locked to the 3.6~THz reference. One can observe a dramatic difference between the two areas, where the DKS-based 300~GHz signal \Erase{comb is made}\Add{becomes} more stable in locked operation. More importantly, we express the fractional instability of the 300~GHz signal in terms of modified Allan deviation (MADEV) in Fig.~\ref{fig4}(b). We choose the MADEV criterion due to the relatively low signal-to-noise ratio of the measured signals, which leads to high white phase noise floor and can hinder the instability level at 1~second averaging time. Free-running operation of the DKS comb at 300~GHz exhibits a stability of $2 \times 10^{-8}$ at 1 second averaging time. Locking the DKS comb to the 300~GHz \Add{domain} drops the stability level to $2 \times 10^{-11}$. This stability level is, in fact, the stability of the 3.6~THz reference signal as optical frequency division has no influence (fractionally) in the spectral purity transfer process.

\section*{Conclusion and outlook}

In summary, we have demonstrated OFD of an optically-carried 3.6~THz reference to the mmW domain with an integrated DKS comb operating at 300~GHz.
\Add{An original measurement scheme is implemented to characterize 300~GHz signal noise with metrological performance, which cannot be accessed through conventional ways relying on the use of a microwave reference.
The measured phase noise levels at 10~kHz are $-100$~dBc/Hz for the generated 300~GHz wave (measurement-limited), and $-108$~dBc/Hz for the repetition rate noise, which are 10~dB and 18~dB lower than any measured noise at 300~GHz, respectively. This leads to attosecond-level absolute timing noise in the mmW frequency range, yet to be explored.} 
The present mmW oscillator, on top of being low noise, has the potential for full integration in a compact fashion. Moreover, \Add{introducing a resonator with a higher \textit{Q} for low power operation of a DKS~\cite{stern2018battery} and} replacing the fiber-based Brillouin cavity with a chip-scale microresonator~\cite{lee2013spiral, gundavarapu2019sub} would lead to a mmW oscillator on chip with spectral purity performance that far outstrips any other chip-based technologies demonstrated to date. 
With a linewidth of $<10$~Hz \Add{estimated from the MADEV}, such an oscillator will impact a wide range of applications ranging from fundamental science to civilian and defense applications where spectral purity, size, weight and energy consumption are critical parameters.
Additionally, reducing the linewidth using rotational spectroscopy~\cite{wang2018chip} could potentially lead to an advanced ultrahigh stability molecular clock on-chip.

\section*{Methods}
\subsection*{Generation of Brillouin Stokes waves}
The two CW lasers pumping the 75-m-long fiber ring cavity are non-resonant with the cavity, instead, the Stokes waves are resonant. This dual-pump consists of a fixed-wavelength external cavity laser diode (RIO, Orion module at 192.4~THz) and a tunable external cavity laser diode (Toptica Photonics, CTL1550 employed at 196~THz). The two CW nodes are combined together through a 50/50 coupler and amplified to an average optical power of 350~mW. The use of only one EDFA is preferable in order to minimize the pump phase noise influence on the Stokes waves. The Stokes wave's phase noise is directly related to the pumps phase noise, so any common noise process will be cancelled out when the two Stokes waves are spatially overlapped and ultimately photodetected. As the free spectral range of the cavity of 2.7~MHz is smaller than a few tens of MHz of Brillouin gain bandwidth, we implement a mode-hopping suppression scheme described in a reference~\cite{danion2016mode} for the laser at 192.4~THz. Exploiting a larger optical frequency span of the DKS is currently limited by available laser wavelengths for the Brillouin pumping and the gain bandwidth of the high power EDFA.

\subsection*{EO comb based 300 GHz}
The EO comb used to generate a 300~GHz wave consists of three phase modulator driven by a dielectric resonant oscillator at 10~GHz. The DRO is split into four paths. One path is reserved for its characterization. The three other paths go through high power RF amplifiers operated at 1~W. The main objective, here, is to generate two stong sidebands separated by 300~GHz. As the phase modulators have a low $V_\pi$ ($\sim 3$~V), an EO comb covering an optical span of $>400$~GHz is possible. Maximizing the optical power of two chosen sidebands with 300~GHz separation is accomplished by finely tuning the DRO frequency. Two optical variable bandpass filters \Add{(Yenista, XTM-50)} spectrally filter the two sidebands of interest. The two sidebands are amplified with an EDFA prior to illumination of the UTC photodiode. Phase-locking of the DRO to the oscillator under test is possible because spectral purity of the DRO is unarguably competitive at a fixed frequency and its modulation bandwidth can be as high as a few hundreds of kHz.

\section*{Acknowledgments}
The authors thank Naoya Kuse and Mark Yeo for their technical contribution at the early stage of this work and Yuzuru Uehara for his help.

\section*{Authors contributions}
T.T. and A.R. conceived the ideas, built, implemented, and operated the experimental setup, and wrote the manuscript. T.N. provided with a UTC-PD and insightful contributions on mmW technology. M.E.F. contributed in the conception of the project and the analysis of the data. T.T. designed and G.N. and M.G. fabricated the SiN microresonator. All the authors contributed to the review of the manuscript. A.R. initiated and supervised the project.

\section*{Competing interests}
The authors declare they have no competing financial interests.

\begin{footnotesize}
\printbibliography[segment=\therefsegment, heading=subbibliography]
\end{footnotesize}

\newpage

\end{refsection}

\clearpage

\setcounter{figure}{0}
\setcounter{section}{0}
\renewcommand{\thefigure}{s\arabic{figure}}
\renewcommand{\thesection}{S\arabic{section}}

\noindent
\begin{center}
\Large{\bf{Supplementary information for \\ ``300~GHz wave with attosecond-level timing noise''}}
\end{center}
\section{Detailed description of experimental methods}
\label{section_setup}
Figure~\ref{figs1} is a schematic drawing of the whole setup for the experiment. There are three sections introduced by gray dashed lines for clarity. 

In section~1, Brillouin Stokes waves of a wavelength tunable laser diode (TLD) at about 1530~nm and a laser diode (LD) 1 at about 1558~nm are generated. Since a Brillouin cavity is located in a different room from most of employed experiment equipment including the TLD, a 10-m-long optical fiber is used to send the TLD light to the cavity. Light from the TLD and the LD1 is amplified with a single high power erbium-doped fiber amplifier (EDFA) up to total power of about 350~mW. For phase noise reduction, a major part of the amplified light is sent into a Brillouin cavity consisting of 75~m optical fiber, a circulator, and an optical coupler (see \ref{section_lasernoise} for more details). The cavity is enclosed in a vacuum chamber and temperature-controlled with 10-mK-level resolution to suppress frequency fluctuation of the optical modes. Back-scattered Brillouin light is generated from the dual-wavelength pump, and coupled with the cavity while the pump light is blocked with the circulator, which allows the presence of the pump inside the fiber all the time regardless of the laser frequency. A proportional-integral-differential (PID) feedback loop is implemented to suppress mode-hopping of the Brillouin light generated from LD1. Small parts of the pump light and Brillouin light are combined and detected at a photodiode (PD) to generate an error signal. The PID loop controls laser frequency of the LD1 to keep the detected Brillouin frequency constant. Output of the Brillouin light is sent to section~2 through a 10~m optical fiber.

Section~2 includes components to control noise in a dissipative Kerr soliton (DKS) comb. Continuous wave (CW) light from LD2 goes through a single sideband modulator (SSBM), an EDFA, and a polarization controller (PC), and is coupled with a silicon nitride (SiN) ring resonator. A DKS comb is initiated by a fast frequency sweep of the input light with the SSBM driven by a control signal from a signal generator (SG) applied through a voltage adder, a voltage control oscillator (VCO), a LNA and a 90 degree hybrid splitter (see \ref{section_microcomb} for more detail). The estimated on-chip pump power is about 250~mW. Polarization of the output light is aligned to the slow axis of a polarization maintaining fiber by a PC and a polarizer (PL) to eliminate need of polarization control in subsequent setups (all employed fibers are polarization maintaining ones except lensed fibers for coupling light with the SiN waveguides and a long fiber delays in section 3). Small portions of the output are used for output power monitoring and an optical spectrum monitor while the remaining large part of the power is split into two by a 60:40 optical coupler. Output from the 60\% port is sent to section~3 for 300~GHz wave generation and characterization, and the light from the 40\% port is injected into a waveshaper (WS). The WS works as a spectrum filter and extract only two comb lines at around 1530~nm and 1558~nm. The comb lines are combined with the Brillouin light generated in section~1 whose power is slightly attenuated by a variable optical attenuator (VOA) to avoid saturation of the PD. The RF output from the PD includes two beat signals; one is originated from the two light at 1530~nm and the other is from the two light at 1558~nm. The RF signals are amplified with a low noise RF amplifier (LNA), filtered with a RF bandpass filter (RF-BPF), and sent to a low-barrier Schottky diode (LBSD) where a beat note of the two beat signals is generated through difference frequency generation. The generated signal is filtered, amplified, divided by 4 with a prescaler and sent to a PID loop filter to generate an error signal for stabilizing DKS comb noise. The obtained error signal is applied to the frequency modulation port of the SSBM besides the control signal for the DKS comb initiation at the voltage adder.

Section~3 demonstrates 300~GHz wave generation and noise measurement. The strong pump light of a DKS comb comb is attenuated with an optical band stop filter (OBSF). The output is amplified with an EDFA and sent to a uni-traveling-carrier photodiode (UTC-PD), where the light given to the UTC-PD is slowly increased by a VOA. With an average photocurrent of 7~mA, a millimeter wave (mmW) with an estimated power of 0.1~mW is generated through photomixing. The mmW is put into the RF port of a fundamental mixer (MX). Two comb lines of an electro-optic (EO) comb are used for generating a mmW signal for the LO port of the MX. The EO comb is obtained with three cascaded phase modulators (PMs) driven by 10~GHz signals from a dielectric resonator oscillator (DRO) amplified by high power amplifiers (HPA). Two lines of the EO comb with separation of 300~GHz are sampled by optical bandpass filters (OBPF) and injected into a UTC-PD, whose output is sent to the MX. Phase noise at offset frequencies of over 50~kHz is measured directly from the MX output. On the other hand, a feedback loop is implemented to lock the DRO noise to the DKS comb noise for the measurement at other offset frequencies since the expected DKS comb noise is lower than the calibrated DRO noise at 300~GHz in the range. The output from the MX is used for generating an error signal, which is applied to the DRO. Then, the locked DRO noise is evaluated either by a signal analyzer (PXA-N9030A, Keysight) or a two-wavelength delayed self-heterodyne interferometer (TWDI) (see \ref{section_TWDI} for more details).
A free-running DKS comb comb noise is evaluated independently with a method shown in \ref{section_300GHzSelf}.

\begin{landscape}
\begin{figure}[htbp]
\hspace{-35pt}
\vspace{100pt}
\includegraphics[width=750pt]{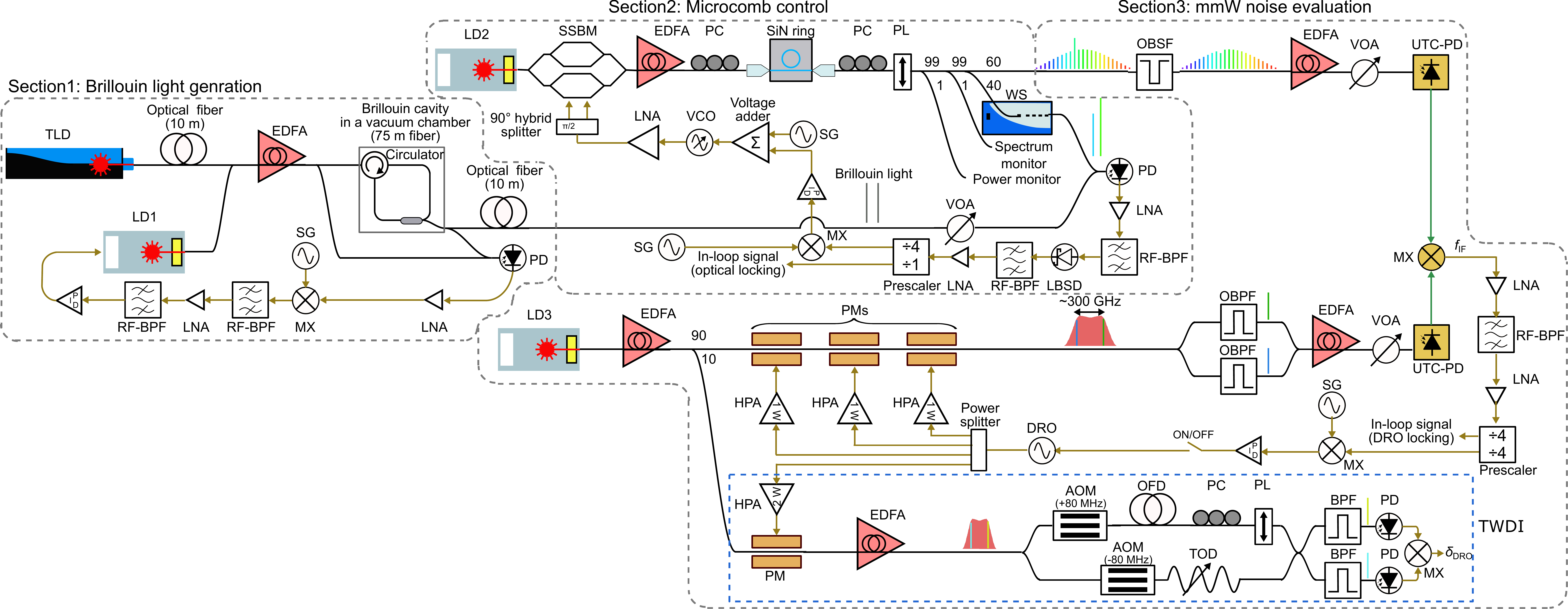}
\caption{A whole picture of the experimental setup for the ultralow noise 300~GHz wave generation and characterization. See text for details.}
\label{figs1}
\end{figure}
\thispagestyle{empty}
\end{landscape}

\section{Laser phase noise reduction through stimulated Brillouin scattering}
\label{section_lasernoise}
Here, effect of phase noise suppression by stimulated Brillouin scattering is described.

In the experiment, light from the LD1 at 1558~nm and the TLD at 1530~nm is non-resonantly injected into the Brillouin cavity. They are strong enough to initiate stimulate Brillouin scattering inside the fiber ring cavity and frequency shifted Brillouin Stokes light is obtained. The Stokes light brings much less phase noise thanks to an ultrahigh $\mathit{Q}$ of the Brillouin cavity and acoustic damping~\cite{debut2000linewidth}. The Brillouin cavity consists of a long fiber of around 75~m, 95:5 optical coupler and a circulator with about 1~dB of insertion loss, which gives an expected optical quality factor of about $8 \times 10^8$ at 1550~nm.

\begin{figure}[htbp]
\centering
\includegraphics[width=400pt]{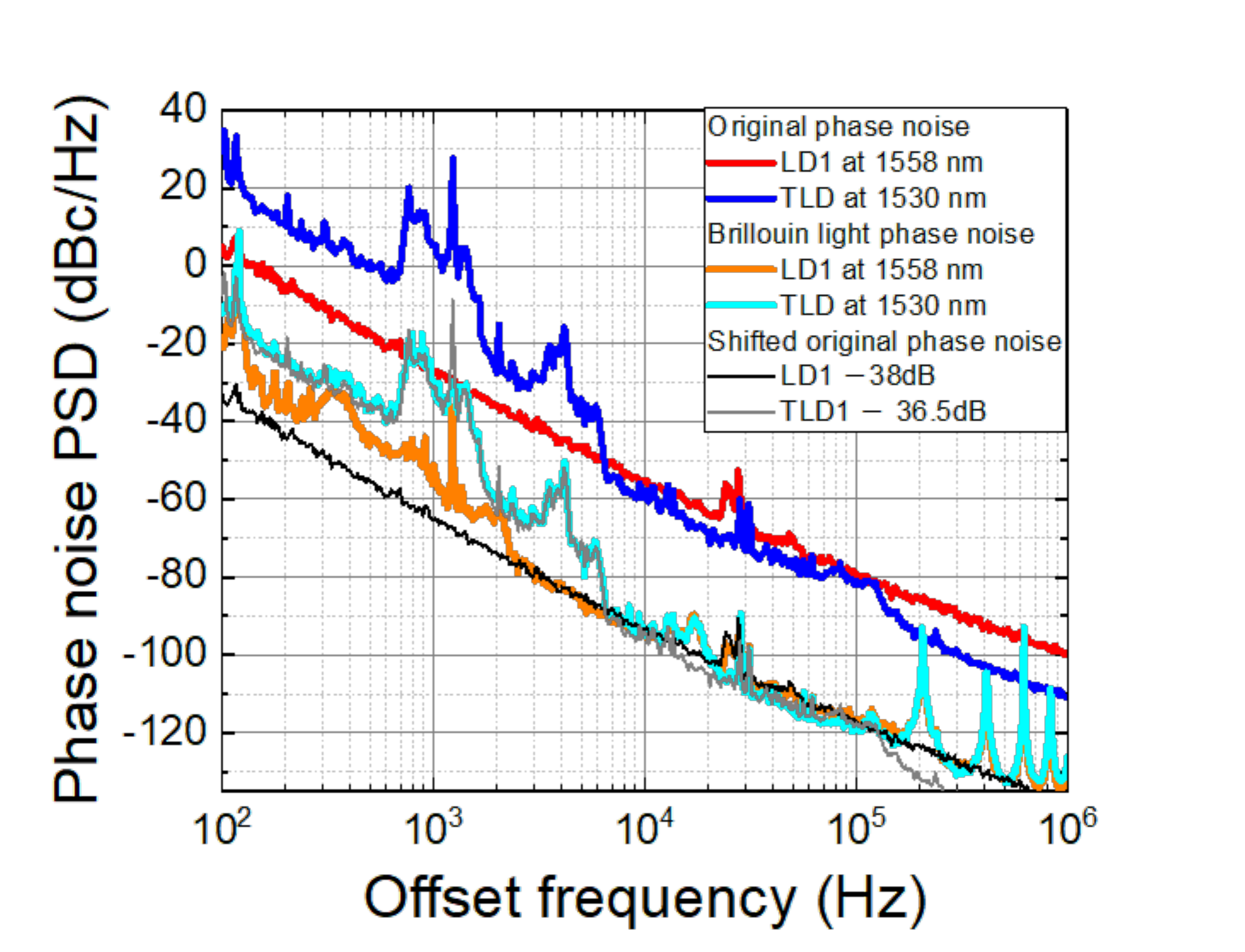}
\caption{Measured phase noise of the two lasers used for the experiment before and after the Brillouin cavity. Employed fiber delay lengths are 10~m for the original phase noise measurement of TLD, 100~m for the original phase noise measurement of LD1, and 1~km for the Brillouin light phase noise measurement. 20~MHz, 2~MHz and 200~kHz are null-frequencies of their transfer functions respectively.}
\label{fig_LaserNoise}
\end{figure}

Laser phase noise is evaluated with a self-heterodyne interferometer with an optical fiber delay (the length of the fiber for each measurement is described in the caption of Fig.~\ref{fig_LaserNoise}). The results are shown in Fig.~\ref{fig_LaserNoise} (red and blue curves are for original phase noise, and orange and light blue curves are for Brillouin Stokes wave noise of LDs at 1558~nm and 1530~nm, respectively). Two curves with specific offsets from the original phase noise are added to the graph for an explanation purpose (black and gray curves). Phase noise reduction of about 38~dB is observed for LD1 noise while a slightly smaller factor of 36.5~dB is obtained for TLD as the black and the grey curves well overlay the orange and the light blue curves, respectively, where the small difference in amount of noise reduction comes from wavelength-dependent losses of the optical components used in the cavity.
Mismatches in noise levels are observed for offset frequencies of 100~Hz to 2.5~kHz for LD1, and 10~kHz to 20~kHz for the both lasers. These are due to the intrinsic noise of the Brillouin cavity.

\newpage

\section{Phase noise measurement with a two-wavelength delayed self-heterodyne interferometer}
\label{section_TWDI}
Here, details of a TWDI denoted by the blue dashed line in Fig.~\ref{figs1} is described. The system is used for characterizing the locked DRO phase noise, the relative phase noise of two Brillouin light, and the DKS comb repetition rate phase noise in the paper.

To evaluate a DRO noise, an EO comb is generated by employing the DRO as a driver. The EO comb is sent to a Mach–Zehnder interferometer (MZI) after amplification by an EDFA up to about 90~mW of total power. The MZI starts from a 50:50 optical coupler. The EO comb on one arm experiences a frequency up-shift by 80~MHz with an acousto-optic modulator (AOM) and time delay given by an optical fiber delay (OFD) followed by polarization alignment with a PC and a PL. On the other arm, a frequency down-shift of 80~MHz is applied to the EO comb and the light passes through a short tunable optical delay (TOD). The comb lines from the two arms are interfered at a 2~x~2 optical coupler and its two outputs are sent to OBPFs and PDs, respectively. Center frequencies of the OBPFs are set at comb line frequencies of $f_1$ and $f_2$ respectively and their mode separation is $m$ (i.e., $f_1 - f_2 = mf_\mathrm{DUT}$, where $f_\mathrm{DUT}$ is DRO frequency under test). The 3~dB bandwidths of the OBPFs are narrower than comb separation of 10~GHz. So, only two lines with frequencies of $f_\mathrm{1\:or\:2} + f_\mathrm{AOM}$ and $f_\mathrm{1\:or\:2} - f_\mathrm{AOM}$ are detected by the PD on each arm, respectively, resulting in generation of beat signals at 160~MHz. The RF signals from the two PDs are mixed at a MX and down-converted to direct current (DC). The DC output gives the following relation \cite{kuse2017electro}:

\begin{equation}
    V_\mathrm{out}(t) \propto \cos (2\pi m f_\mathrm{DUT} \tau +m(\phi _\mathrm{DUT} (t) -\phi _\mathrm{DUT} (t-\tau)))
\label{eqs1}
\end{equation}
where $\phi _\mathrm{DUT}$ is phase noise of the DRO under test and $\tau$ is time delay given by the long OFD. When a quadrature condition of $2\pi mf_\mathrm{DUT}\tau=\pi/2$ is satisfied, phase noise PSD of the DRO of $L_\mathrm{DUT}(f)$ for offset frequency $f$ relates to the voltage noise PSD of $V_\mathrm{out}(f)$ by,
\begin{equation}
    V_\mathrm{out}(f)^2 = K_1 \cdot m^2 \cdot |H(if)|^2 \cdot L_\mathrm{DUT}(f)
\end{equation}
\begin{center}
or
\end{center}
\begin{equation}
L_\mathrm{DUT}(f) =  \frac{V_\mathrm{out}(f)^2}{K_1 \cdot m^2 \cdot |H(if)|^2}.
\end{equation}

\begin{figure}[htbp]
\centering
\includegraphics[width=\linewidth]{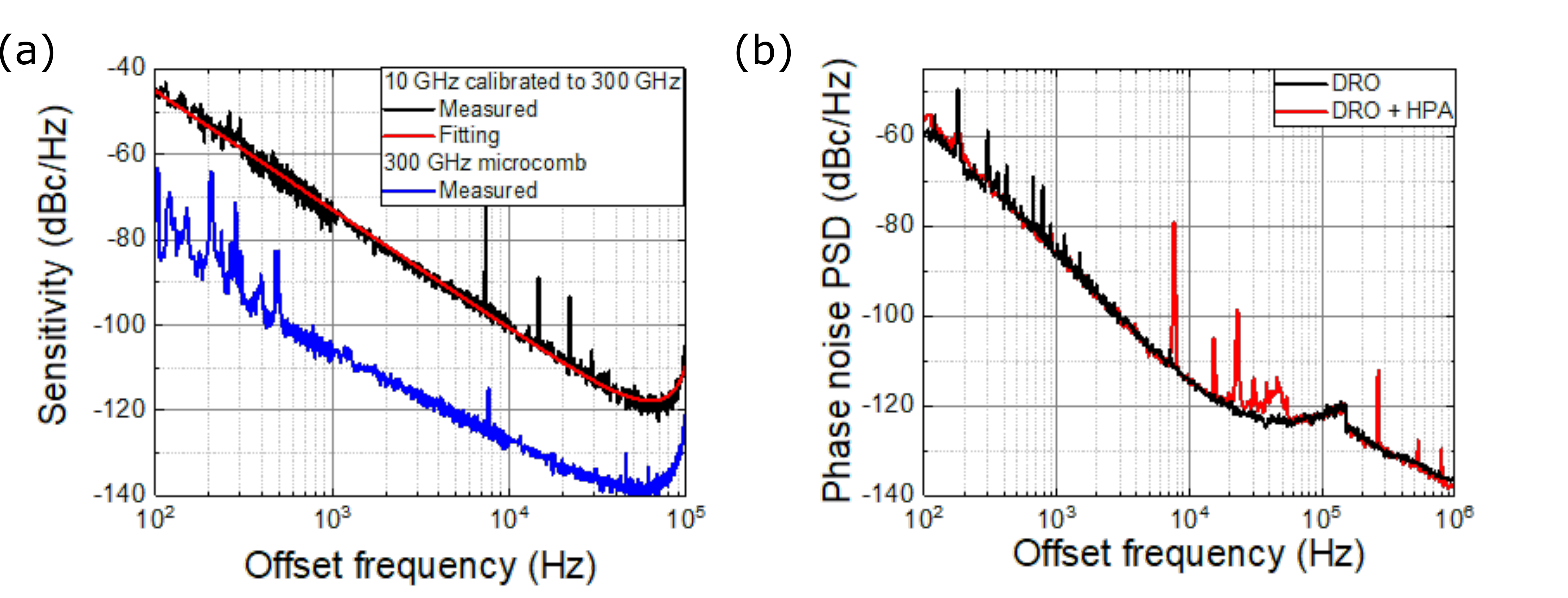}
\caption{(a) Sensitivity of the TWDI. The OFD length is about 2~km, and the employed mode separations are $m=20$ for the EO comb based 10~GHz wave noise measurement and $m=10$ for the direct DKS comb noise measurement. (b) DRO phase noise with and without a HPA.}
\label{fig_TWDI}
\end{figure}

$K_1$ is a coefficient to include all gain/loss in the system except the division factor $m^2$ (e.g., carrier power to convert dBm/Hz into dBc/Hz, conversion gain/loss at a mixer) and $H(if)$ is a delay transfer function. 
In the experiment, $K_1$ is derived from a measurement of known noises of an oscillator or a peak in a spectrum. The quadrature condition is ensured by controlling the TOD and monitored during the measurement. Noise floor of the system is evaluated through a measurement having a same mode on the both PDs. The measured sensitivity and its fitting are shown in Fig.~\ref{fig_TWDI}(a) where only the fitting curve is presented in the main text. 

For the characterization of the relative phase noise of the two Brillouin light and the phase noise of a DKS comb, their own optical lines are used for the measurement instead of using ones from an EO comb. Since the sensitivity of the method is not limited by a carrier frequency of interest, the method can have an effectively lower noise floor for a comb with a larger frequency separation when it is compared at a specific calibrated carrier frequency. As an example, an estimated sensitivity limit for a direct 300~GHz DKS comb noise measurement is added to Fig.~\ref{fig_TWDI}(a) where its sensitivity is about 34~dB lower at 1~kHz than that for 10~GHz wave calibrated to 300~GHz (the difference is mainly originated from a division factor of $20\mathrm{log}10(30)=29.5$~dB and 6~dB coming from the employed mode separation difference).
Another technical noise limitation is a HPA. As shown in Fig.~\ref{fig_TWDI}(b), one of the HPAs in the system adds spurious peaks to the DRO noise, which can be observed in the measurement results.

\newpage

\section{Phase noise measurement of 300~GHz wave with a photonic self-heterodyne delayed interferometer}
\label{section_300GHzSelf}
\begin{figure}[htbp]
\centering
\includegraphics[width=450pt]{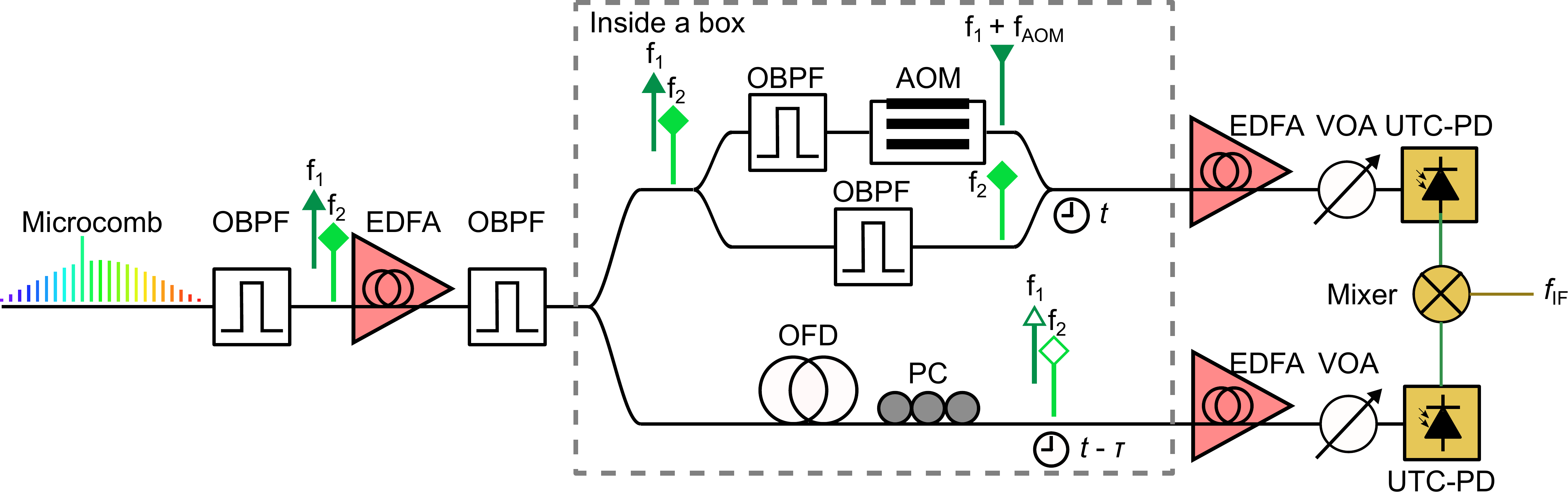}
\caption{Schematic of a developed photonic self-heterodyne delayed interferometer for free-running 300~GHz wave noise measurement.}
\label{fig_300GHzSelf}
\end{figure}

Here, a measurement system used for characterizing a free-running DKS comb noise is described (see Fig.~\ref{fig_300GHzSelf} for its schematic). 

Adjacent two lines of a DKS comb with 300~GHz FSR are sampled with an OBPF with a wide bandwidth of $>300$~GHz, and amplified with an EDFA. Another wide-band OBPF is used to eliminate amplified spontaneous emission noise of the EDFA after that. The output is sent to a 50:50 optical coupler where one port is for shifting frequency separation of the two lines and the other port is for adding time delay. 
On the upper arm for the frequency shift, the path is split into two again with an optical coupler, and the two lines are filtered with OBPFs so that different lines go through different arms. Since only one line experiences frequency shift with an AOM, the frequency separation of the two lines is shifted by the AOM frequency when they are combined.
From the output of the lower arm of the first coupler, time-delayed two lines are obtained owing to a long OFD. A PC is used to compensate polarization rotation in the OFD.
The two outputs are sent to different UTC-PDs respectively after power amplification and control with EDFAs and VOAs. Generated 300~GHz waves are interfered at a mixer, and an IF signal with the AOM frequency is obtained. The phase noise of the IF signal is expressed by,
\begin{eqnarray}
 \delta f_{\mathrm{IF}}(t) &=& (\delta f_{1}(t) + \delta f_{\mathrm{AOM}}(t) - \delta f_{2}(t)) - (\delta f_{1}(t-\tau) - \delta f_{2}(t-\tau)) \nonumber \\
&=& \delta f_{\mathrm{300G}}(t) - \delta f_{\mathrm{300G}}(t-\tau) + \delta f_{\mathrm{AOM}}(t),
\end{eqnarray}
where $\delta f_{\mathrm{IF}}$, $\delta f_{1}$, $\delta f_{2}$, $\delta f_{\mathrm{AOM}}$, $\delta f_{\mathrm{300G}}$ are phase noise of the IF signal, two comb lines, the AOM, and the generated 300~GHz wave. Since the AOM noise is negligibly small compared to 300~GHz wave noise, the IF noise is presented in frequency domain as,
\begin{equation}
    L_\mathrm{IF}(f) = L_\mathrm{300G}(f) |H(if)|^2,
\end{equation}
where $f$, $L_\mathrm{IF}$, $L_\mathrm{300G}(f)$, $|H(if)|^2$ are offset frequency from the carrier frequency, phase noise power spectrum density of the IF signal and 300~GHz wave, and transfer function of the OFD. So, 300~GHz wave noise is obtained from the IF noise and the OFD transfer function. 200~m OFD is used in the experiment. 

Though this method worked for the free-running DKS comb noise measurement, the sensitivity was not high enough for its locking condition due to obtained high white noise floor of -98~dBc/Hz limited by small signal-to-noise of the optical lines at the mixer (the sensitivity corresponds to around -74~dBc/Hz at 10~kHz offset with 200~m OFD).

\section{Properties of a generated DKS comb}
\label{section_microcomb}
Here, features of a generated DKS comb is described to show its consistent nature with dissipative Kerr solitons in previous studies.  

\begin{figure}[htbp]
\centering
\includegraphics[width=\linewidth]{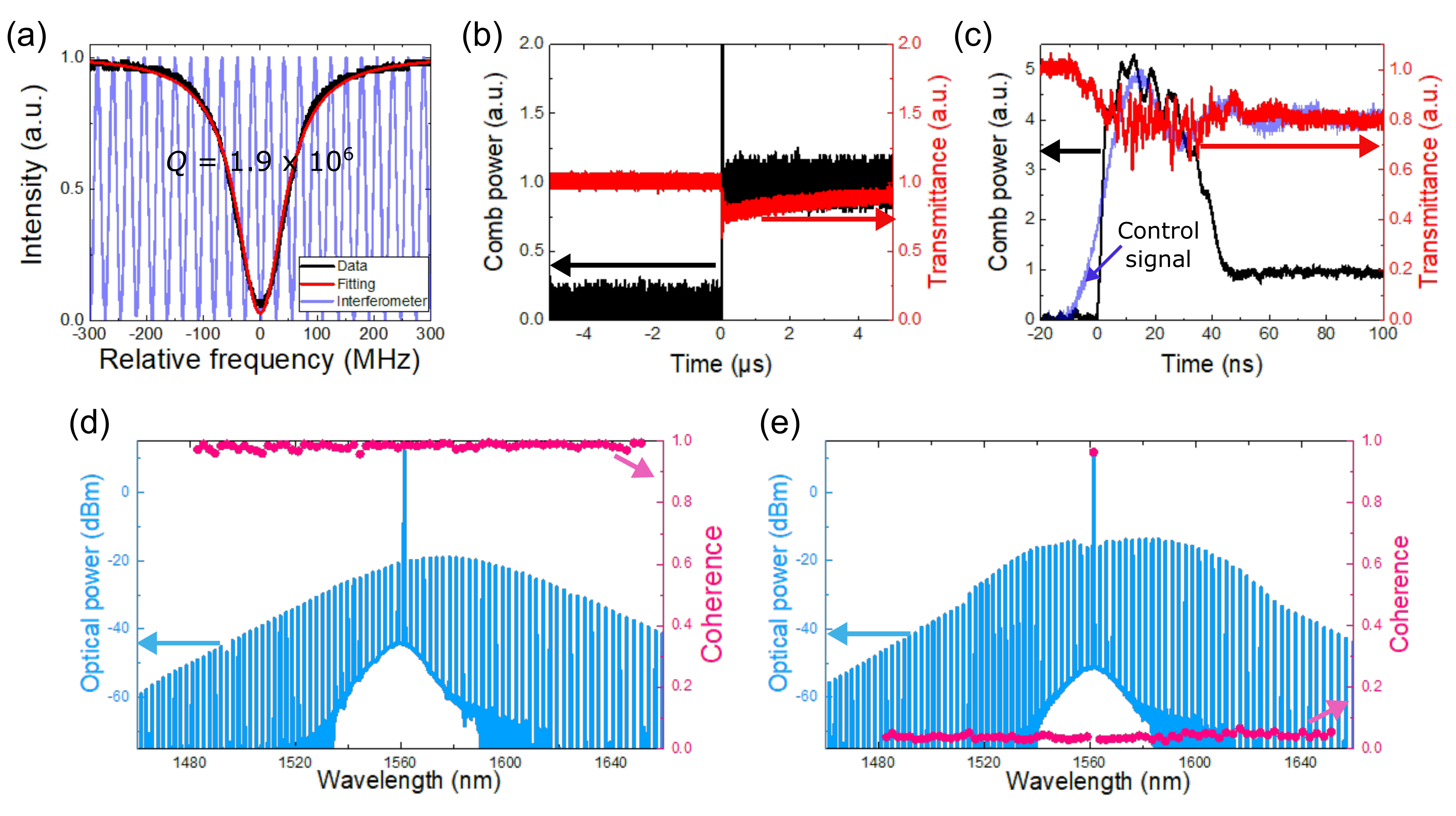}
\caption{(a) Transmission spectrum of a resonant mode. The scan frequency is calibrated with a MZI with a frequency period of about 27~MHz (blue line). (b) Observed soliton step. Black and red curves show comb power excluding the pump and transmitted power including the pump, respectively. (c) Enlarged view of (b). Blue shaded curve shows a waveform of a control signal in an arbitrary unit. Timing difference of each signal is roughly compensated by measuring lengths of optical fibers and electrical cables. Optical spectrum and coherence of generated (d) single soliton and (e) MI combs.}
\label{fig_microcomb}
\end{figure}

The resonant mode of the SiN ring resonator used for the DKS comb generation is shown in Fig.~\ref{fig_microcomb}(a). The obtained loaded $\mathit{Q}$ is around $1.9 \times 10^6$ and the mode is close to a critical coupling condition. 
A DKS comb is launched by a fast sweep of a pump frequency by about 2~GHz with a SSBM. Comb power excluding the strong pump and transmittance including the pump power are measured during the scanning process and shown in Fig.~\ref{fig_microcomb}(b) and (c) to see power transitions known as soliton steps \cite{herr2014temporal}. The comb power goes up and the transmittance drops almost instantly when a control signal is applied to the SSBM at time of $\sim$ 0~\textmu s as shown in Fig.~\ref{fig_microcomb}(b). After the formation of the comb, transmittance gradually increases owing to slow thermal response of the resonator, however, the small power change does not impact on the presence of the generated comb \cite{PhysRevApplied.14.014006}. More microscopic transition is shown in Fig.~\ref{fig_microcomb}(c). As the control signal rises and the pump frequency is red-tuned (blue shaded line), the transmission starts decreasing, and the comb power shows up at around 0~ns. At first, the transmission and the comb power fluctuate relatively largely, which implies that the comb is in noisy modulation instability (MI) state. Overshoot of the control signal gives corresponding small power fluctuation, and a transition to a stable soliton state is triggered during it ($>45$~ns).
Though the number of transition steps can be more than two, the final comb state represented by Fig.~\ref{fig_microcomb}(d) is well reproducible for specific scan parameters.

Next, coherence of each comb line of two states of Kerr microresonator frequency combs (a DKS comb and a MI comb) is measured by following the method described in \cite{webb2016measurement, kim2019turn}. A fiber-based asymmetric MZI interferometer is employed for the measurement, where its path-length difference of 17~m corresponds to about 52 photon lifetimes of the resonant mode. 
Intensities of each comb line is measured with an optical spectrum analyzer (OSA) under a condition that the MZI is exposed to air current, which allows the MZI to take all possible phases thanks to the introduced instability. 100 OSA traces are used for the coherence calculation. Visibility is calculated for each comb line from maximum and minimum intensities obtained in the traces.
Since two optical couplers in the MZI are connected so that their wavelength-dependence is canceled out each other and almost same amount of power come from each arm at the interference, the calculated visibility is assumed to be equal to first-order coherence.
The measured coherence is shown in Fig.~\ref{fig_microcomb}(d) and (e). The average high coherence of 0.98 is obtained for a single soliton comb. On the other hand, coherence is low over entire range except the pump light, and the average is about 0.05 for a MI comb generated through a slow frequency scanning. This difference is expected for Kerr frequency combs in the different states of MI and cavity soliton \cite{erkintalo2014coherence}.
Also, observed small envelop's peak shift from the pump frequency in the soliton spectrum is probably attributed to Raman self-frequency shift observed for a soliton comb in \cite{karpov2016raman}.

\clearpage

\begin{footnotesize}
\printbibliography[segment=\therefsegment, heading=subbibliography]
\end{footnotesize}

\end{document}